\documentclass[aps,nofootinbib,prd,superscriptaddress,tightenlines,notitlepage,twocolumn,showpacs]{revtex4}

\usepackage[colorlinks]{hyperref}
\usepackage{tabularx}
\usepackage{graphicx}
\usepackage{amssymb,bm}
\usepackage{xcolor}
\usepackage{amsmath}
\usepackage[varg]{txfonts}
\usepackage{enumerate}
\usepackage[normalem]{ulem}

\newcommand{\apjs}{Astrophys. J. Suppl. Ser.}
\newcommand{\aap}{Astron. \& Astrophys.}

\newcommand{\mnras}{Mon. Not. R. Astron. Soc.}

\newcommand{\pasp}{Pub. Astro. Soc. Pacific}

\newcommand{\ssr}{Space Sci. Rev.}

\newcommand{\jgr}{J. Geophys. Res.}
\newcommand{\cjaa}{Chin. J. Astron. Astrophys.}

\newcommand{\PMO}{\affiliation{Purple Mountain Observatory, Chinese Academy of Sciences, Nanjing 210023, China}}
\newcommand{\USTC}{\affiliation{School of Astronomy and Space Sciences, University of Science and Technology of China, Hefei 230026, China}}

\begin{document}

\title{Scale invariance in X-ray flares of gamma-ray bursts}

\author{Jun-Jie Wei}\thanks{jjwei@pmo.ac.cn}\PMO\USTC

\begin{abstract}
X-ray flares are generally believed to be produced by the reactivation of the central engine, and
may have the same energy dissipation mechanism as the prompt emission of gamma-ray bursts (GRBs).
X-ray flares can therefore provide important clues to understanding the nature of the central engines
of GRBs. In this work, we study for the first time the physical connection between differential size
and return distributions of X-ray flares of GRBs with known redshifts. We find that the differential
distributions of duration, energy, and waiting time can be well fitted by a power-law function.
In particular, the distributions for the differences of durations, energies, and waiting times
at different times (i.e., the return distributions) well follow a $q$-Gaussian form. The $q$ values
in the $q$-Gaussian distributions remain nearly steady for different temporal interval scales,
implying a scale-invariant structure of GRB X-ray flares. Moreover, we verify that the $q$ parameters
are related to the power-law indices $\alpha$ of the differential size distributions, characterized as
$q=(\alpha+2)/\alpha$. These statistical features can be well explained within the physical framework
of a self-organizing criticality system.
\end{abstract}



\maketitle

\section{Introduction}
X-ray flares are one of the most intriguing phenomena in the afterglow phase of gamma-ray bursts
(GRBs) in the \emph{Swift} era \citep{2005Sci...309.1833B,2006ApJ...642..389N,2006ApJ...642..354Z,
2007ApJ...671.1903C,2007ApJ...671.1921F,2010MNRAS.406.2113C,2011MNRAS.417.2144M,2011A&A...526A..27B}.
Nearly one third of \emph{Swift} GRBs have remarkable X-ray flares with rapid rise and fall times
\citep{2007ApJ...671.1903C,2007ApJ...671.1921F}. Unexpected X-ray flares with an isotropic-equivalent
energy release from $10^{48}$ to $10^{53}$ ergs have been observed in both long and short GRBs
\citep{2006A&A...454..113C,2006ApJ...641.1010F,2006A&A...450...59R,2011MNRAS.417.2144M}. They
typically occur at a few seconds to $10^{6}$ seconds after the GRB trigger \citep{2011A&A...526A..27B}.
By studying the temporal behaviors and spectral properties, there is a general agreement that X-ray
flares are produced by late central engine activities, and may share the same physical origin with
the prompt emission of GRBs \citep{2005Sci...309.1833B,2006RPPh...69.2259M,2007ChJAA...7....1Z}.

The concept of self-organized criticality (SOC) \citep{1986JGR....9110412K,1987PhRvL..59..381B}
has been widely applied to explain the dynamical behaviors of astrophysics systems
\citep{2016SSRv..198...47A}. A generalized definition of SOC is a critical state of a nonlinear
dissipation system that is continuously and slowly driven towards a critical value of a system-wide
instability threshold, resulting in scale-free, fractal-diffusive, and intermittent avalanches
with power-law-like size distributions \citep{2014ApJ...782...54A}. Thus, the emergence of
scale-free power-law size distributions is the fundamental property that SOC systems have
in common \citep{2011soca.book.....A,2012A&A...539A...2A,2015ApJ...814...19A}. Interestingly,
Ref.~\cite{2013NatPh...9..465W} compared the statistical properties of 83 GRB X-ray flares and
11,597 solar hard X-ray flares from \emph{RHESSI} during 2002-2007. They found that
the differential distributions of energy and duration for GRB X-ray flares exhibit power-law tails
similar to those of solar flares, but with different power-law indices. Although the power-law indices
for both classes of flares are apparently different, Ref.~\cite{2013NatPh...9..465W} suggested that
these statistical properties can be well understood with the framework of fractal-diffusive SOC
systems \citep{footnote1}. The solar-flare-like behavior also indicates that GRB X-ray flares
may be driven by magnetic reconnection processes \citep{footnote2}. Some theoretical models
have indeed argued that GRB X-ray flares could be magnetically dominated explosive events
\citep{2006Sci...311.1127D,2011MNRAS.413.2031M,2006MNRAS.370L..61P}.

It was proposed that another SOC hallmark is the scale invariance of the avalanche size differences
\citep{2007PhRvE..75e5101C,2015EPJB...88..206W}. Ref.~\cite{2007PhRvE..75e5101C} showed that when
criticality appears, the probability density functions (PDFs) for the avalanche size differences at
different times have the shape of $q$-Gaussian. Moreover, such a $q$-Gaussian form does not depend on
the time interval adopted for the avalanche size difference, and it is rightly so when considering
energy differences between real earthquakes \citep{2007PhRvE..75e5101C}. That is, the $q$ values
in $q$-Gaussian distributions are nearly equal for different scale intervals, implying a
scale-invariant structure of earthquakes (see also \cite{2015EPJB...88..206W}).
Ref.~\cite{2017ChPhC..41f5104C} found that the soft gamma repeater (SGR) J1550--5418 has the property
of scale invariance similar to that of earthquakes. Ref.~\cite{2021ApJ...920..153W} further confirmed
that the PDFs of the differences of fluences, fluxes, and durations for other SGRs (i.e., SGR 1806--20
and SGR J1935+2154) also exhibit a common $q$-Gaussian distribution at different scale intervals,
which indicates that there is a common scale-invariant property in SGRs (see also
\cite{2022MNRAS.510.1801S}). Ref.~\cite{2020MNRAS.491.2156L} analyzed 93 bursts from the repeating
fast radio burst (FRB) 121102 in a continuous observation by the Green Bank Telescope, and showed
that FRB 121102 has a similar scale-invariant behavior. Very recently, the detection of 1652
independent bursts from FRB 121102 using the Five-hundred-meter Aperture Spherical radio Telescope
has been reported \citep{2021Natur.598..267L}. Based on this largest burst sample,
Ref.~\cite{2021ApJ...920..153W} suggested that scale invariance in both FRB 121102 and SGRs can be
explained with the physical framework of fractal-diffusive SOC systems.

Besides the power-law distribution of the avalanche size, the $q$-Gaussian distribution of
the avalanche size differences at different times (i.e., the scale-invariant structure) provides
a new powerful way for characterizing the presence of SOC \citep{2007PhRvE..75e5101C}.
Although power-law size distributions have been found in GRB X-ray flares \citep{2013NatPh...9..465W},
their scale-invariant property has not yet been explored. On the other hand, a theoretical relation between
the power-law index $\alpha$ of the avalanche size distribution and the $q$ value of the appropriate
$q$-Gaussian has been proposed \citep{2007PhRvE..75e5101C,2010PhRvE..82b1124C}. Taking advantage of
this relation, one can easily estimate $q$ parameter values of the appropriate $q$-Gaussians a priori
from the known power-law indices $\alpha$ of the system. In this work, we study for the first time
the scale-invariant behaviors of GRB X-ray flares, and then test the validity of the proposed relation
between the values of $q$ and $\alpha$ using real data.

The rest of the paper is arranged as follows. The differential distributions of duration, energy, and
waiting time of GRB X-ray flares are shown in Section~\ref{sec:PL}. In Section~\ref{sec:scale}, for the
first time, we calculate the cumulative distributions of the differences of these three
quantities and investigate the scale-invariant property. The evidence that may shed light on the theoretical
$q$--$\alpha$ relation is presented in Section~\ref{sec:relation}. Lastly, discussions and conclusions
are given in Section~\ref{sec:summary}.

\section{Power-law Size Distributions of GRB X-ray Flares}
\label{sec:PL}
Recently, Ref.~\cite{2016ApJS..224...20Y} presented a catalog of GRB X-ray flares observed
by \emph{Swift} between 2005 April and 2015 March. A total of 200 X-ray flares of GRBs with known
redshifts are reported in this catalog, representing the largest sample at present. The catalog
contains the start time $t_{\rm start}$, end time $t_{\rm end}$, and fluence $F$ of each flare.
In this work, we use this sample to study the statistical properties of the energy release, duration,
and waiting-time distributions of GRB X-ray flares. The isotropic energy of each flare in the
0.3--10 keV energy band can be calculated by $E_{\rm iso}=4\pi D_{\rm L}^{2}(z)F/(1+z)$, where
$z$ is the redshift and $D_{\rm L}(z)$ is the luminosity distance calculated for a flat $\Lambda$CDM
cosmological model with $H_{0}=67.36$ km s$^{-1}$ Mpc$^{-1}$, $\Omega_{\rm m}=0.315$, and
$\Omega_{\Lambda}=0.685$ \citep{2020A&A...641A...6P}. The flare duration in the source frame
can be obtained by $T=(t_{\rm end}-t_{\rm start})/(1+z)$, where $(1+z)$ accounts for the relativistic
time dilation factor. The waiting time in the source frame can be estimated by
$\Delta t=(t_{{\rm start},i+1}-t_{{\rm start},i})/(1+z)$, where $t_{{\rm start},i+1}$
and $t_{{\rm start},i}$ are the observed start times of the $(i+1)$-th and $i$-th flares,
respectively. For the first flare appearing in a GRB, the source-frame waiting time is taken to be
$t_{{\rm start},1}/(1+z)$.

\begin{figure*}
\begin{center}
\vskip-0.1in
\includegraphics[width=0.33\textwidth]{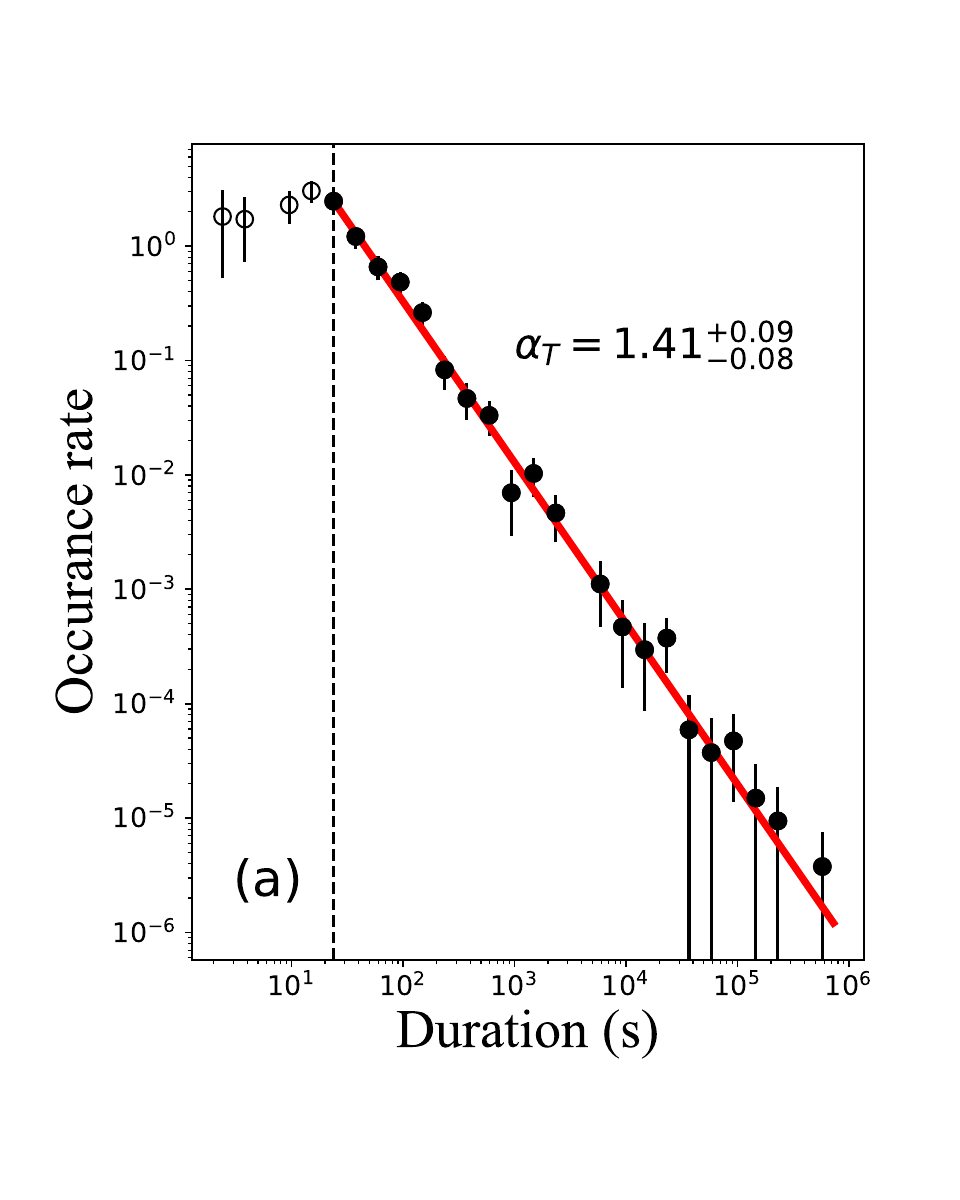}
\includegraphics[width=0.33\textwidth]{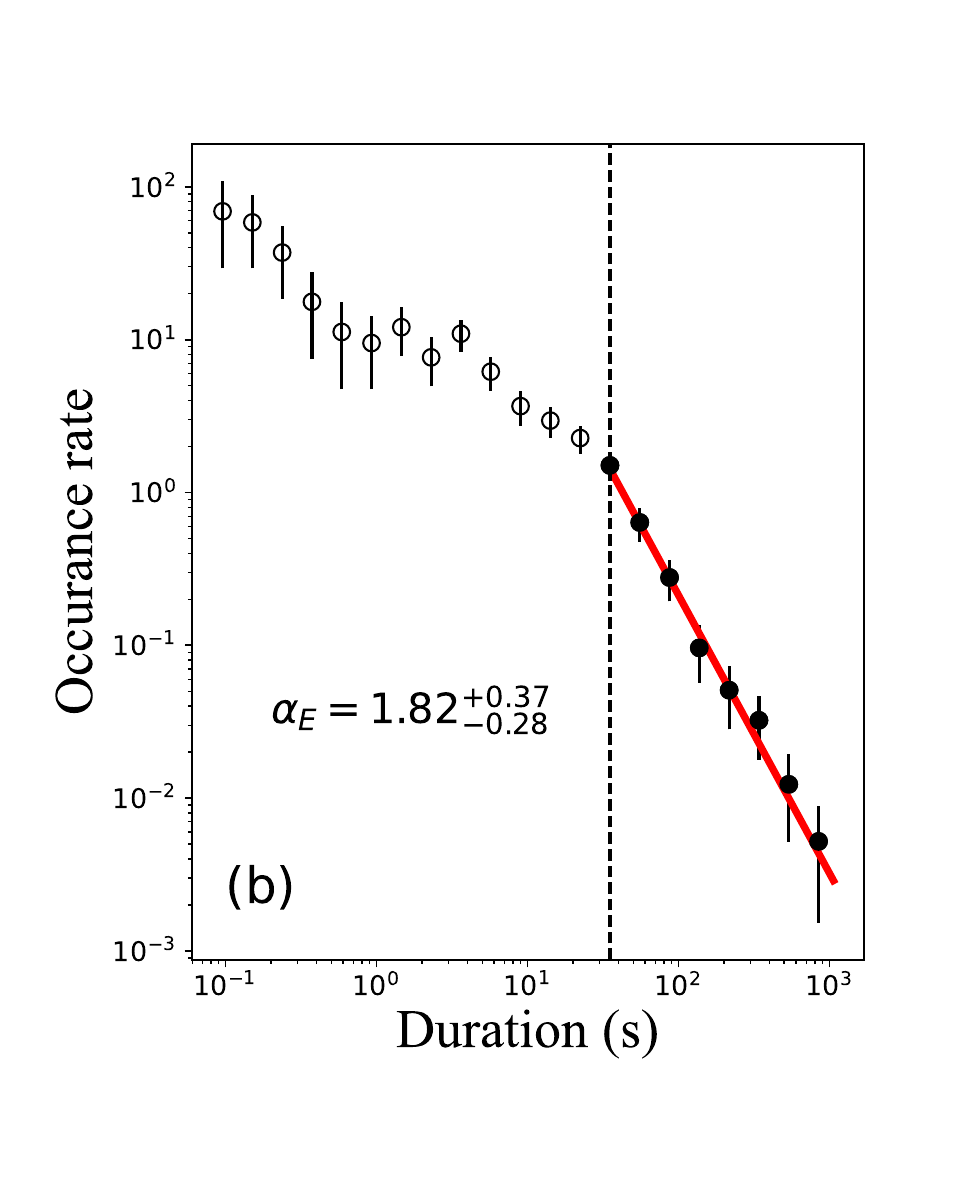}
\includegraphics[width=0.33\textwidth]{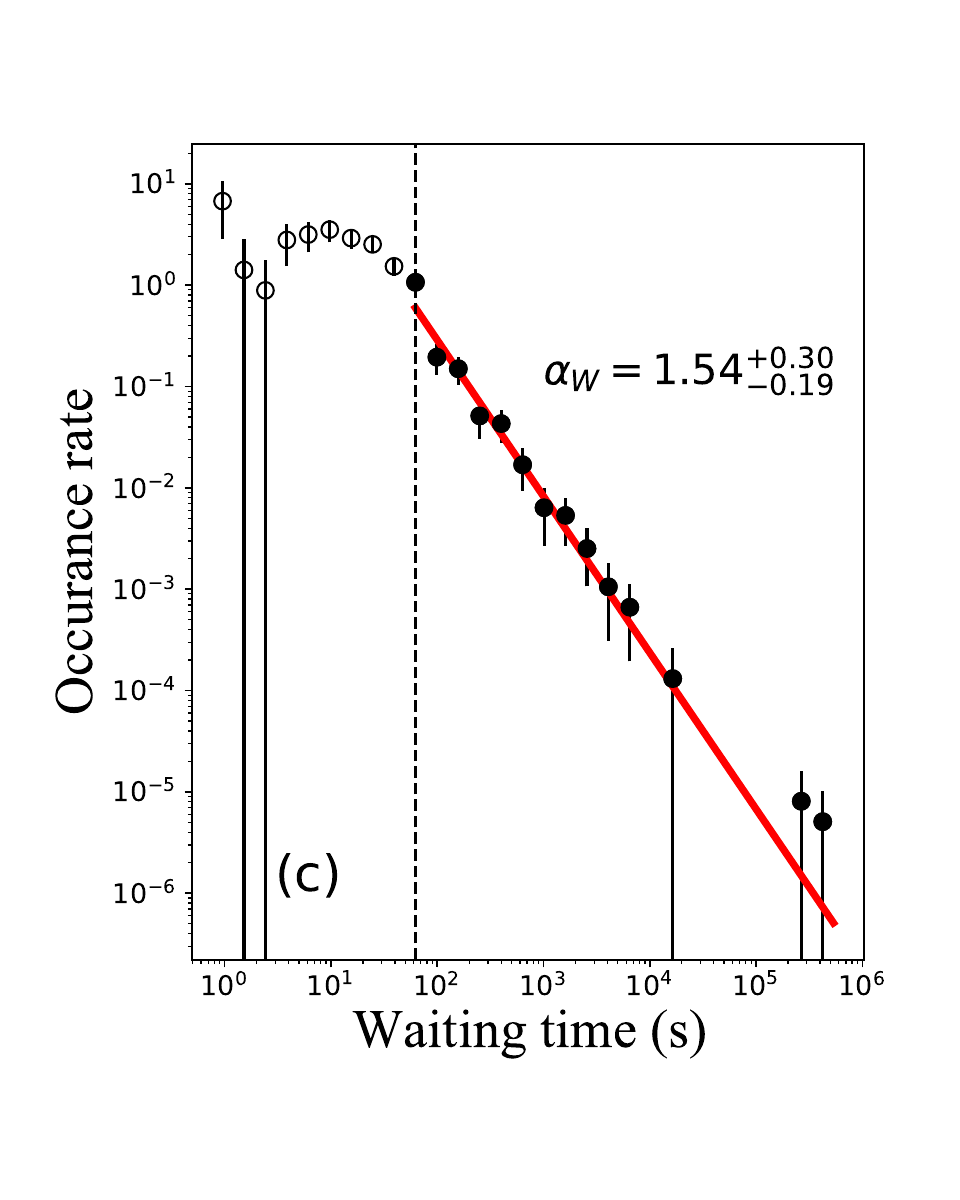}
\vskip-0.4in
\caption{Differential distributions of duration, energy, and waiting time for GRB X-ray flares.
The red solid lines are the best-fitting results of power laws. Only the filled dots above the
thresholds (vertical dashed lines) are fitted.}
\label{fig:f1}
\vskip-0.2in
\end{center}
\end{figure*}

Power-law-like size distributions are ubiquitous in astrophysical instabilities
\citep{2015ApJ...814...19A}. That is, the differential size distribution $N_{\rm diff}(s)$ of
extreme events can be well described with a power-law function,
\begin{equation}
N_{\rm diff}(s){\rm d}s\propto s^{-\alpha_{s}}{\rm d}s\;,
\label{eq:pl}
\end{equation}
where $\alpha_{s}$ is the power-law index. Here we explore the differential distributions of
duration, energy, and waiting time for GRB X-ray flares, and then fit them with the power-law
model. We apply a uniformly logarithmic binning to the data for the differential distributions.
Empirically, the number of bins is set to be $\log_{10}(s_{2}/s_{1})\times5$, where $s_1$ and
$s_2$ are the minimum and maximum values of the sampled data, respectively
\citep{2015ApJ...814...19A}. The counted number of events per (logarithmic) bin is $N_{{\rm bin},i}$,
while the resulting differential distribution is determined by dividing with the (non-equidistant)
bin width $\Delta s_{i}$, so that $N_{{\rm diff}}(s)=N_{{\rm bin},i}/\Delta s_{i}$.
The expected uncertainty of the differential distribution is then
\begin{equation}
\sigma_{{\rm diff},i}=\sqrt{N_{{\rm bin},i}}/\Delta s_{i}\;.
\end{equation}
As shown in Figure~\ref{fig:f1}, due to incomplete sampling of the smallest events, the observed
differential distribution often deviates from an ideal power law at the lower end. It is therefore
necessary to find a lower bound or threshold $s_{0}$ that defines an upper range [$s_0$, $s_2$]
where events are completely sampled and a power-law fit can be performed. A reliable threshold
$s_0$ can be obtained from the bin where the counted number of events per bin has a maximum, i.e.,
${\rm max}\left[N_{\rm bin}(s_i)\right]=N_{\rm bin}(s_{i}=s_{0})$ \citep{2015ApJ...814...19A}.
The determined threshold values $s_0$ of the differential distributions of duration, energy, and
waiting time are $23.84$ s, $3.51\times10^{51}$ erg, and $62.46$ s, respectively (corresponding
to the vertical dashed lines in Figure~\ref{fig:f1}).

\begin{table}
\renewcommand\arraystretch{1.5}
\tabcolsep=0.4cm
\centering \caption{The best-fitting power-law indices $\alpha$ of differential size distributions
and the average $q$ values of the appropriate $q$-Gaussians for GRB X-ray flares}
\begin{tabular}{cccc}
\hline
\hline
 Parameters  &  Duration  &  Energy  &  Waiting time  \\
\hline
$\alpha$     &  $1.41^{+0.09}_{-0.08}$   &  $1.82^{+0.37}_{-0.28}$  &  $1.54^{+0.30}_{-0.19}$ \\
$q$          &  $2.57^{+0.04}_{-0.04}$   &  $2.22^{+0.10}_{-0.10}$  &  $2.37^{+0.04}_{-0.04}$ \\
\hline
\end{tabular}
\label{table1}
\end{table}

For the upper range [$s_0$, $s_2$], we obtain the best-fitting parameters by minimizing the
$\chi^{2}$ statistic,
\begin{equation}
\chi^{2}=\sum_{i}\frac{\left[N_{{\rm diff},i}^{\rm obs}-N_{{\rm diff}}^{\rm fit}(s_{i})\right]^{2}}{\sigma_{{\rm diff},i}^{2}}\;,
\end{equation}
where $N_{{\rm diff},i}^{\rm obs}$ is the observed differential distribution and
$N_{{\rm diff}}^{\rm fit}(s_{i})$ is the theoretical distribution function obtained from
Equation~(\ref{eq:pl}). We adopt the Python implementation, EMCEE \citep{2013PASP..125..306F},
to apply the Markov Chain Monte Carlo technique to perform the fitting. From Figure~\ref{fig:f1},
we find that the differential distributions (above the threshold $s_0$) of duration, energy, and
waiting time for GRB X-ray flares can be well fitted by a power-law function (see red solid lines).
The best-fitting parameters and their $2\sigma$ uncertainties are presented in Table~\ref{table1}.
The power-law indices of duration, energy, and waiting time are $\alpha_{T}=1.41^{+0.09}_{-0.08}$
(with a reduced $\chi_{\rm dof}^{2}=0.54$), $\alpha_{E}=1.82^{+0.37}_{-0.28}$
($\chi_{\rm dof}^{2}=0.17$), and $\alpha_{W}=1.54^{+0.30}_{-0.19}$
($\chi_{\rm dof}^{2}=0.97$), respectively (all quoted uncertainties will hereafter
be at the 95\% confidence level). These power-law distributions are natural predictions of
SOC theory \citep{2011soca.book.....A}.

\begin{table*}
\renewcommand\arraystretch{1.3}
\tabcolsep=0.5cm
\centering \caption{Fit results and estimated 95\% confidence level constraints on the $q$ values for $n=1$, 20, and 40}
\begin{tabular}{lcccc}
\hline
\hline
 & $n$  &  1  &  20  &  40  \\
\hline
Duration & $q$ &  $2.53^{+0.04}_{-0.04}$   &  $2.57^{+0.04}_{-0.04}$  &  $2.61^{+0.05}_{-0.05}$ \\
         & $\chi^{2}/{\rm d.o.f.}$  &  $45.66/196=0.23$   &  $17.07/177=0.10$  &  $14.61/157=0.09$ \\
\hline
Energy & $q$ &  $2.15^{+0.06}_{-0.07}$   &  $2.25^{+0.07}_{-0.07}$  &  $2.27^{+0.08}_{-0.08}$ \\
         & $\chi^{2}/{\rm d.o.f.}$  &  $18.94/196=0.10$   &  $29.19/177=0.16$  &  $12.04/157=0.08$ \\
\hline
Waiting time & $q$ &  $2.37^{+0.04}_{-0.04}$   &  $2.39^{+0.04}_{-0.04}$  &  $2.35^{+0.05}_{-0.05}$ \\
         & $\chi^{2}/{\rm d.o.f.}$  &  $24.70/196=0.13$   &  $15.04/177=0.08$  &  $13.99/157=0.09$ \\
\hline
\end{tabular}
\label{table2}
\end{table*}

\begin{figure*}
\begin{center}
\vskip-0.1in
\includegraphics[width=0.45\textwidth]{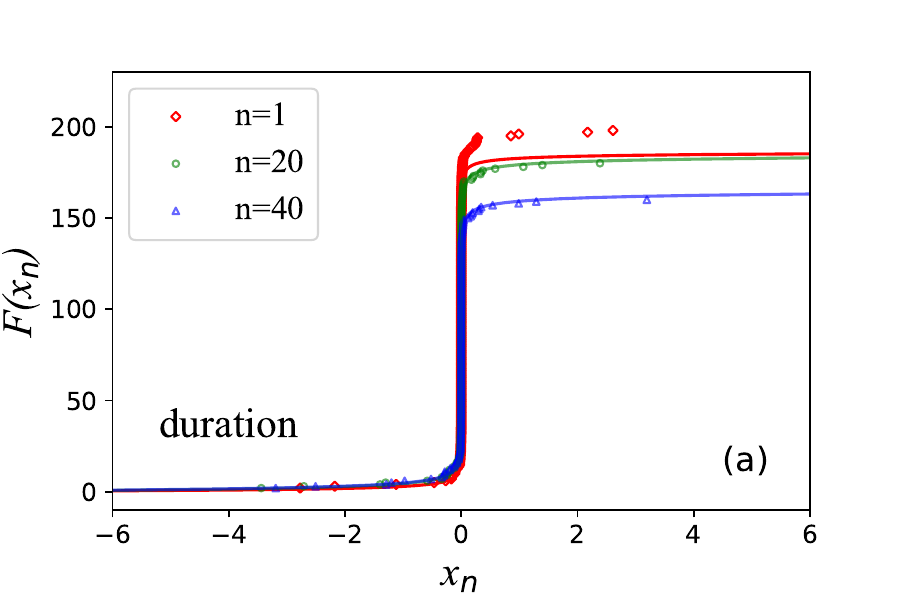}
\includegraphics[width=0.45\textwidth]{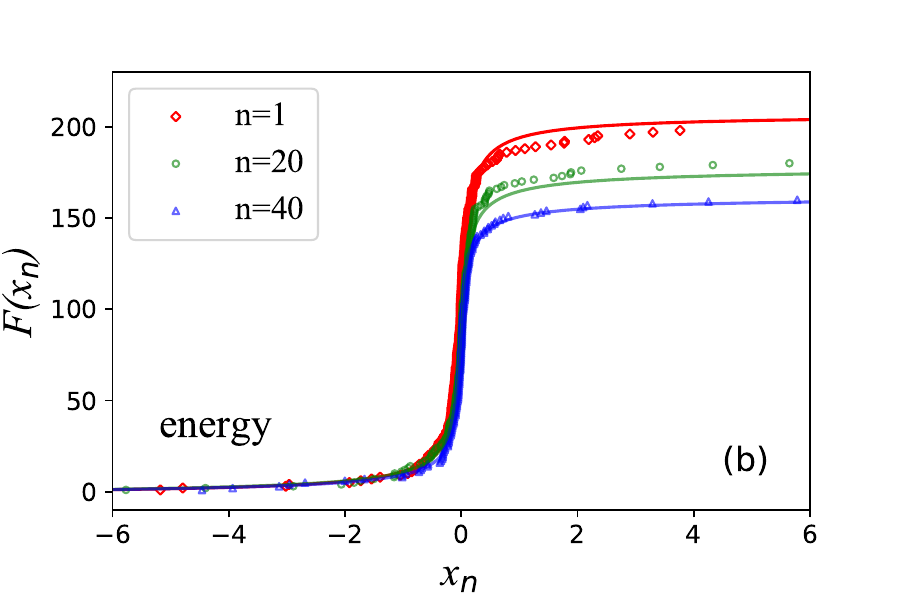}
\includegraphics[width=0.45\textwidth]{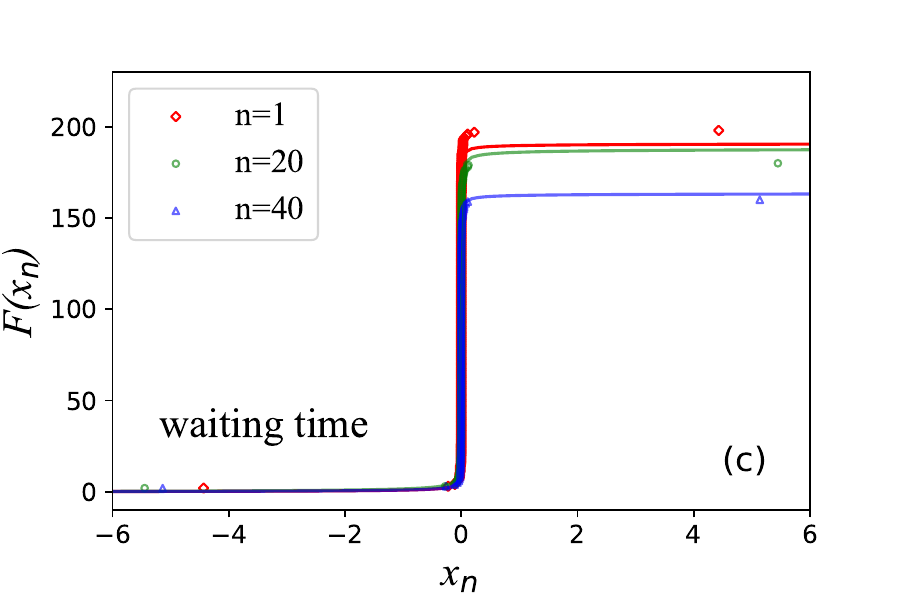}
\includegraphics[width=0.45\textwidth]{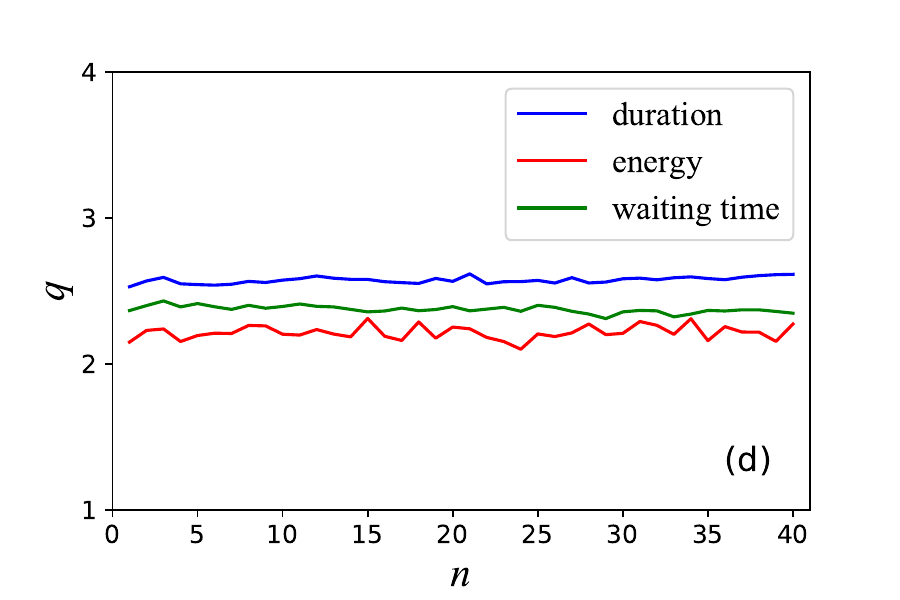}
\caption{Scale-invariant structure of GRB X-ray flares. Cumulative distributions of the differences
of durations (panel (a)), energies (panel (b)), and waiting times (panel (c)) for $n=1$ (diamonds),
$n=20$ (circles), and $n=40$ (triangles), and the best-fitting $q$-Gaussian distributions
(solid lines). Panel (d): the best-fitting $q$ values in the $q$-Gaussian distribution as
a function of $n$.}
\label{fig:f2}
\vskip-0.2in
\end{center}
\end{figure*}

\section{Scale Invariance in GRB X-ray Flares}
\label{sec:scale}
SOC models have been widely investigated considering time intervals between avalanches in the
critical regime \citep{2004PhRvL..92j8501C}. In this section, we focus our attention on the
``returns'' $X_{n}=s_{i+n}-s_{i}$, i.e., on the differences between two avalanche sizes, where
$s_{i}$ is the size (duration, energy, or waiting time) of the $i$-th flare in temporal order,
and the integer $n$ is the temporal interval scale. In practice, $X_{n}$ is rescaled as
$x_{n}=X_{n}/\sigma_{X_{n}}$, where $\sigma_{X_{n}}$ is the standard deviation of $X_{n}$.
Here we study the statistical distributions of the dimensionless returns $x_{n}$.

The return distributions of earthquakes, SGRs, and repeating FRB 121102 have been shown to follow
the $q$-Gaussian function
\citep{2007PhRvE..75e5101C,2015EPJB...88..206W,2017ChPhC..41f5104C,2020MNRAS.491.2156L,2021ApJ...920..153W,2022MNRAS.510.1801S}.
The $q$-Gaussian function is defined as \citep{1988JSP....52..479T,1998PhyA..261..534T}
\begin{equation}
f(x_{n})=A\left[1-B\left(1-q\right)x_{n}^{2}\right]^{\frac{1}{1-q}}\;,
\label{eq:q-Gauss}
\end{equation}
where $A$ is a normalization constant, and the parameters $B$ and $q$ determine the width and sharpness
of the distribution, respectively.
When $q\rightarrow1$, the $q$-Gaussian function reduces to an ordinary Gaussian function
with $\mu=0$ and $\sigma=1/\sqrt{2B}$. Thus, $q\neq1$ indicates a deviation from Gaussian statistics.
Inspired by this, we try to fit the return distributions of GRB X-ray flares using the $q$-Gaussian
function. Due to the limited number of GRB X-ray flares, we use the cumulative distribution of
$q$-Gaussian in the fitting to avoid binning,
\begin{equation}
F(x_{n})=\int_{-\infty}^{x_{n}}f(x_{n}){\rm d}x_{n}\;.
\label{eq:CDF}
\end{equation}
For a fixed $n$, the free parameters ($A$, $B$, and $q$) can be optimized by minimizing
the $\chi^{2}$ statistics,
\begin{equation}
\chi^{2}=\sum_{i}\frac{\left[N_{\rm cum}(<x_{n,i})-F(x_{n,i})\right]^{2}}{\sigma_{{\rm cum},i}^{2}}\;,
\end{equation}
where $\sigma_{{\rm cum},i}=\sqrt{N_{\rm cum}(<x_{n,i})}$ is the uncertainty of the data point,
with $N_{\rm cum}(<x_{n,i})$ being the cumulative number of the returns. In our analysis, we choose
wide flat priors for $\log_{10}A\in[-3,\;8]$, $\log_{10}B\in[-3,\;8]$, and $q\in(1,\;1000]$.

Figure~\ref{fig:f2} displays some examples of the fits for the data of GRB X-ray flares.
In this figure, we show the cumulative distributions of the differences of durations (panel (a)), energies
(panel (b)), and waiting times (panel (c)) for the temporal interval scale $n=1$ (diamonds),
$n=20$ (circles), and $n=40$ (triangles). To make the data points to be more distinguishable,
their uncertainties are not plotted in this figure. But the uncertainties have been employed
to weight the data points in the fitting procedure. Note that for a fixed $n$, the number
of the dimensionless returns $x_{n}$ should be $(200-n)$, where $200$ is the total number of
the avalanche sizes. That is, the maximum values of the data points in Figure~\ref{fig:f2}
are $199$, $180$, and $160$ for $n=1$, $n=20$, and $n=40$, respectively. The red, green,
and blue smooth curves show the $q$-Gaussian fits for $n=1$, $n=20$, and $n=40$, respectively.
If the normalized cumulative distributions of the size differences at different temporal intervals
are considered, then we would see that the data points in each panel of Figure~\ref{fig:f2}
are almost independent of $n$, and that the three fitted curves overlap
almost completely. The best-fitting $q$ values and their $2\sigma$ uncertainties for $n=1$,
$20$, $40$ are listed in Table~\ref{table2}, along with the $\chi^{2}$ value for the fit.
One can see that the cumulative distributions of the differences of durations, energies,
and waiting times are well reproduced by means of $q$-Gaussians.

Furthermore, we calculate the cumulative distributions of the differences of durations, energies,
and waiting times at different scale intervals $1\leq n \leq40$, and fit the cumulative distributions
with the $q$-Gaussian function. Figure~\ref{fig:f2}(d) shows the best-fitting $q$ values as a function
of $n$. We find that the $q$ values are nearly invariant and independent of $n$. The mean values of
$q$ for duration, energy, and waiting time are $q_{T}=2.57\pm0.04$, $q_{E}=2.22\pm0.10$, and
$q_{W}=2.37\pm0.04$, respectively, which are listed in Table~\ref{table1}. Here the uncertainties
represent the $2\sigma$ standard deviations of $q$ values. The steadiness of $q$ values for different
temporal interval scales $n$ implies the scale invariance of GRB X-ray flares. This property is
very similar to that of earthquakes \citep{2007PhRvE..75e5101C,2015EPJB...88..206W}, SGRs
\citep{2017ChPhC..41f5104C,2021ApJ...920..153W,2022MNRAS.510.1801S}, and repeating FRBs
\citep{2020MNRAS.491.2156L,2021ApJ...920..153W}.

\begin{figure}
\begin{center}
\vskip-0.2in
\includegraphics[width=0.5\textwidth]{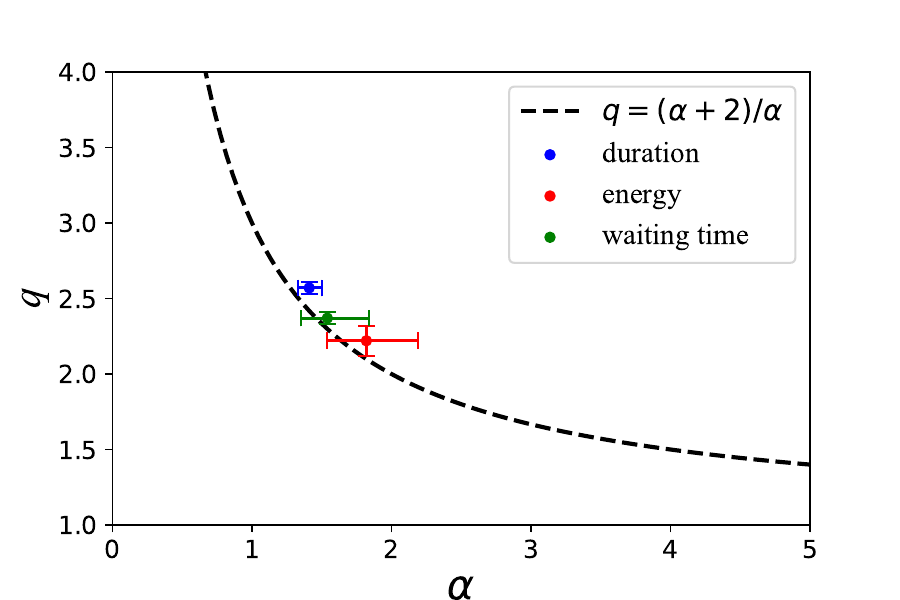}
\caption{The relationship between the values of $q$ and $\alpha$. The dashed line shows the expected relation
from Equation~(\ref{eq:q-alpha}). Data points with $2\sigma$ error bars correspond to the real values of
$q$ and $\alpha$ found for the durations (blue dot), energies (red dot), and waiting times (green dot) of
GRB X-ray flares.}
\label{fig:f3}
\vskip-0.2in
\end{center}
\end{figure}

\section{Avalanche Size and Return Distributions}
\label{sec:relation}
As mentioned above, Ref.~\cite{2007PhRvE..75e5101C} proposed an analysis method to characterize
the presence of SOC in the limited number of earthquakes by taking advantage of the return
distributions [i.e., distributions of the size (released energy) differences of the avalanches
(earthquakes) at different times]. They presented the first evidence that the return distributions
are in the shape of $q$-Gaussians, standard distributions generating naturally within the framework
of non-extensive statistical mechanics \citep{1988JSP....52..479T,1998PhyA..261..534T}.
In the hypothesis of no correlation between the sizes of two avalanches, a theoretical relation
connecting the power-law index $\alpha$ of the avalanche size distribution and the $q$ value of
the appropriate $q$-Gaussian has been derived as \citep{2007PhRvE..75e5101C,2010PhRvE..82b1124C}
\begin{equation}
q=\frac{\alpha+2}{\alpha}\;,
\label{eq:q-alpha}
\end{equation}
which is rather important because it enables the $q$ parameter determined a priori from the
well-known power-law index $\alpha$ of the system. The $q$--$\alpha$ relation expected from
Equation~(\ref{eq:q-alpha}) is shown in Figure~\ref{fig:f3} (dashed line). It is obvious that
the $q$ value decreases with increasing $\alpha$, and $q$ tends to $1$ when $\alpha \rightarrow \infty$.

With the best-fitting power-law indices $\alpha$ of differential size distributions and the average
$q$ values of the appropriate $q$-Gaussians for GRB X-ray flares in hand, we now have the opportunity
to test the validity of Equation~(\ref{eq:q-alpha}). The observed $\alpha$-$q$ values for duration,
energy, and waiting time are plotted as the blue, red, and green dots in Figure~\ref{fig:f3}, along
with their corresponding $2\sigma$ error bars. The values we get for the avalanche size power
laws are $\alpha^{\rm obs}_{T}=1.41^{+0.09}_{-0.08}$, $\alpha^{\rm obs}_{E}=1.82^{+0.37}_{-0.28}$,
and $\alpha^{\rm obs}_{W}=1.54^{+0.30}_{-0.19}$, which correspond, according to
Equation~(\ref{eq:q-alpha}), to the theoretical values of $q^{\rm theo}_{T}=2.42^{+0.09}_{-0.08}$,
$q^{\rm theo}_{E}=2.10^{+0.22}_{-0.17}$, and $q^{\rm theo}_{W}=2.30^{+0.25}_{-0.16}$, respectively.
The observed $q$ values are $q^{\rm obs}_{T}=2.57\pm0.04$, $q^{\rm obs}_{E}=2.22\pm0.10$,
and $q^{\rm obs}_{W}=2.37\pm0.04$, which agree well with the predictions of Equation~(\ref{eq:q-alpha})
at the confidence levels of $3.0\sigma$, $1.0\sigma$, and $0.6\sigma$, respectively.
These results support the validity of the theoretical relation (Equation~(\ref{eq:q-alpha}))
to a certain extent.

\section{Summary and Discussion}
\label{sec:summary}
SOC phenomena can be identified and diagnosed by analyzing the power-law or power-law-like
distributions of the avalanche sizes, and therefore the determination of the power-law index
$\alpha$ is an important  quest. Alternative, real SOC dynamics can also be discriminated in
a quantitative way by analyzing avalanche size differences. When the system is approaching to
a critical state, the PDFs for the avalanche size differences at different times are well fitted
with a $q$-Gaussian function. Another remarkable feature is the so-called scale invariance,
where the $q$ values in the $q$-Gaussian distributions are nearly invariant and independent of
the temporal interval scale adopted. Moreover, a theoretical relation between the power-law index
$\alpha$ and the $q$ value of the appropriate $q$-Gaussian has been established as
$q=(\alpha+2)/\alpha$ \citep{2007PhRvE..75e5101C,2010PhRvE..82b1124C}. This is important as
it makes the $q$ parameter estimated a priori and therefore it rescues $q$ from being a fitting
parameter.

In this work, we investigate the statistical properties of X-ray flares of GRBs with known redshifts.
We first show that the differential distributions of duration, energy, and waiting time of GRB X-ray
flares can be well fitted by a power-law function, confirming previous findings. We then find that
the probability distributions of the differences of durations, energies, and waiting times can be well
fitted by a $q$-Gaussian function and the $q$ values remain nearly constant for different time scales.
Our research reveals for the first time that GRB X-ray flares have the scale-invariant property.
Lastly, we verify that the real values of $q$ and $\alpha$ found for the durations, energies, and
waiting times of GRB X-ray flares are roughly consistent with the predictions of
Equation~(\ref{eq:q-alpha}). This is the first study to test the validity of Equation~(\ref{eq:q-alpha})
using real data. Our findings support the argument that a dissipative SOC mechanism with long-range
interactions is at the origin of GRB X-ray flares. In the future, many more X-ray flares of GRBs
will be detected. The physical connection between avalanche size and return distributions can be
further investigated based on the larger sample of GRB X-ray flares.

\begin{acknowledgments}
This work is partially supported by the National Key Research and Development
Program of China (2022SKA0130100), the National Natural Science Foundation of China
(grant Nos.~11725314 and 12041306), the Key Research Program of Frontier Sciences
(grant No. ZDBS-LY-7014) of Chinese Academy of Sciences,
the Natural Science Foundation of Jiangsu Province (Grant No. BK20221562),
and the Young Elite Scientists Sponsorship Program of Jiangsu Association for Science and Technology.
\end{acknowledgments}


\end{document}